\documentclass[printer]{aa}
\usepackage{psfig}

\def\ltsima{$\; \buildrel < \over \sim \;$}
\def\lsim{\lower.5ex\hbox{\ltsima}}
\def\gtsima{$\; \buildrel > \over \sim \;$}
\def\gsim{\lower.5ex\hbox{\gtsima}}

\begin{document}

\title{The optical afterglow of GRB~000911: evidence for an
associated supernova?\thanks{Based on observations made with ESO
Telescopes at the Paranal Observatories under programme IDs 65.H--0215
and 266.D--5620. Some of the data presented here were obtained at the
W.  M.  Keck Observatory, which is operated as a scientific
partnership among the California Institute of Technology, the
University of California, and the National Aeronautics and Space
Administration.  The Observatory was made possible by the generous
financial support of the W. M. Keck Foundation.}}

\titlerunning{Supernova in GRB~000911}
\authorrunning{Lazzati et al.}

\author{D. Lazzati\inst{1}, S. Covino\inst{2}, G. Ghisellini\inst{2},
D. Fugazza\inst{2}, S. Campana\inst{2}, P. Saracco\inst{2},
P. A. Price\inst{3,4}, E. Berger\inst{3}, S. Kulkarni\inst{3}, 
E. Ramirez--Ruiz\inst{1}, A. Cimatti\inst{5},
M. Della Valle\inst{5}, S. di Serego Alighieri\inst{5},  
A. Celotti\inst{6}, F. Haardt\inst{7}, G. L. Israel\inst{8} \and 
L. Stella\inst{8}}

\institute{
Institute of Astronomy, University of Cambridge, Madingley Road,
CB3 0HA Cambridge, England
\and
Osservatorio Astronomico di Brera, Via Bianchi 46, I--23807
Merate (Lc), Italy
\and
Palomar Observatory, 105-24, California Institute of
Technology, Pasadena, CA, 91125
\and
Research School of Astronomy \& Astrophysics, Mount
Stromlo Observatory, Cotter Road, Weston, ACT, 2611, Australia
\and
Osservatorio Astrofisico di Arcetri, Largo E. Fermi 5, I--50125 Firenze, Italy
\and
SISSA/ISAS, via Beirut 4, I--34014 Trieste, Italy
\and
Universit\`a dell'Insubria, Via Lucini 3, I--22100 Como, Italy
\and
Osservatorio Astronomico di Roma, Via Frascati 33, I--00040 Monteporzio 
Catone, Italy
}

\offprints{D. Lazzati, \\
\email{lazzati@ast.cam.ac.uk}}

\abstract{We present photometric and spectroscopic observations of the 
late afterglow of GRB~000911,  starting $\sim 1$~days after the burst
event and lasting $\sim 8$~weeks.  We  detect a moderately significant
re--brightening in the $R$, $I$ and $J$ lightcurves, associated with a
sizable reddening of the spectrum.   This can be explained through the
presence of  an underlying supernova,  outshining the  afterglow $\sim
30$~days   after   the  burst   event.  Alternative explanations   are
discussed.
}

\maketitle

\section{Introduction}
\label{sec:int}

An anomalous re--brightening has been detected in the optical
lightcurves of at least two gamma--ray bursts $\sim 30$ days after the
burst event (GRB~980326: Bloom et al. 1999; GRB~970228: Reichart 1999,
Galama et al.  2000).  These have been tentatively interpreted as due
to the simultaneous explosion of a supernova (SN) with a lightcurve
similar to that of SN1998bw (Galama et al. 1998) which outshines the
afterglow emission at the time of the SN peak. GRB~980326 showed a
clear excess in $R$ band (aided by having a very faint host galaxy)
and the spectrum changed from blue (early) to red close to the peak of
rebrightening; however, neither the redshift of the afterglow nor the
host is known to date (Bloom et al.  1999). GRB~970228 had the
advantage of a known redshift (Djorgovskij et al.  1999), a moderate
amount of multiband data but lacked spectroscopic coverge.  A broad
band multifilter spectral energy distribution (SED) of GRB~970228 at
day $\sim 30$ was extracted (Reichart 1999), but the $K$ band
measurement, important to constrain the reddening, was questioned by
Galama et al. (2000).

The  confirmation  of a   GRB--SN   connection is clearly   of primary
importance to  solve the problem of  the burst  progenitor (Eichler et
al.  1989; Woosley  1993; Paczy\`nski 1998),   but  also bears important
consequences  on  the  $\gamma$--ray radiation mechanism   (Lazzati et
al. 2000)  and the origin of emission  and absorption features  in the
X--ray spectra  of  the burst proper  and   its afterglow (Lazzati  et
al. 1999; Amati et al. 2000; Piro et al.  2000; Rees \& Meszaros 2000;
Vietri et al. 2001; Lazzati et al. 2001).

Here  we   present  the   results   of  a  simultaneous  multi--filter
observational campaign designed at detecting and studying the spectrum
of  the re--brightening component  in the burst  of September 11$^{\rm
th}$, 2000. We observed the optical transient (OT)  in five filters at
three epochs. A low resolution spectrum was taken $\sim 36$ days after
the burst explosion.

GRB~000911   was detected  by Ulysses,   NEAR  and Konus  on September
11.30237  (Hurley 2000).    It had  a  duration   of $\sim 500$~s,   a
25--100~keV fluence of ${\cal  F}\sim 5 \times  10^{-6}$~erg~cm$^{-2}$
and         a    peak             flux         over   0.5~s         of
$\sim9\times10^{-7}$~erg~cm$^{-2}$~s$^{-1}$.    Its  afterglow     was
detected  at radio wavelengths  (8.46~GHz,  Berger \& Frail 2000)  and
subsequently confirmed in the optical  $R$ band (Berger  et al.  2000)
at  a position of R.A.  $02^{\rm   h}18^{\rm m}34^{\rm s}.36$ and Dec.
$07^\circ44'27''.7$  (Price  2000).  The OT   displayed a fairly rapid
decay law $F_\nu(t)  \propto  t^{-1.4}$ (Price  et al. 2000,  see also
Sect.~\ref{sec:modpho})  in the $R$   band.  The redshift  of the host
galaxy was measured to be $z = 1.06$ (Price et al. 2001).

\section{Observations}
\label{sec:obs}

We observed  the OT associated with GRB  000911  with the ESO/VLT-Antu
telescope  (instruments,  FORS1   and ISAAC),  the   Keck I  telescope
(instrument, LRIS) and  the MSO 50-inch  telescope; see Tables 1 and 2
for log  of observations.    The  VLT observations were  designed   to
optimize the chances of detecting a re-brightening due to a SN similar
to that of SN 1998bw. Furthermore,  we obtained low resolution spectra
with the FORS1  instrument (grism 150I,  blocking filter OG590) on day
36, around the time of the expected peak SN emission.

\subsection{Photometry}
\label{sec:obspho}

All scientific frames were reduced in a standard way with the
ESO--MIDAS package (98NOV and 99NOV versions).  Photometric analysis
was carried out with DAOPHOT\,II (Stetson 1987) as implemented in the
ESO--MIDAS and IRAF (version 2.11.3) packages.  We derived both
aperture and point--spread function (PSF) fitting photometry.  The
model PSF was always based on a large number of stars.  Seeing
conditions were good in the first and third VLT runs, ($\sim
0.8$\,arcsec), and exceptional in the second one ($\sim 0.5$\,arcsec).
For all three VLT runs suitable standard star fields were observed:
\object{SA98}, \object{Mark\,A} and \object{SA113} in the optical;
\object{FS28}, \object{S234} and \object{P530} in the $J$ band. The
observations were all performed under photometric conditions.  A
comparison of our photometric calibration and that by Henden (2000)
shows that ours is slightly fainter ($\sim 0.1$\,mag) for stars at the
faint end of Henden's list.

\begin{figure*}
\psfig{file=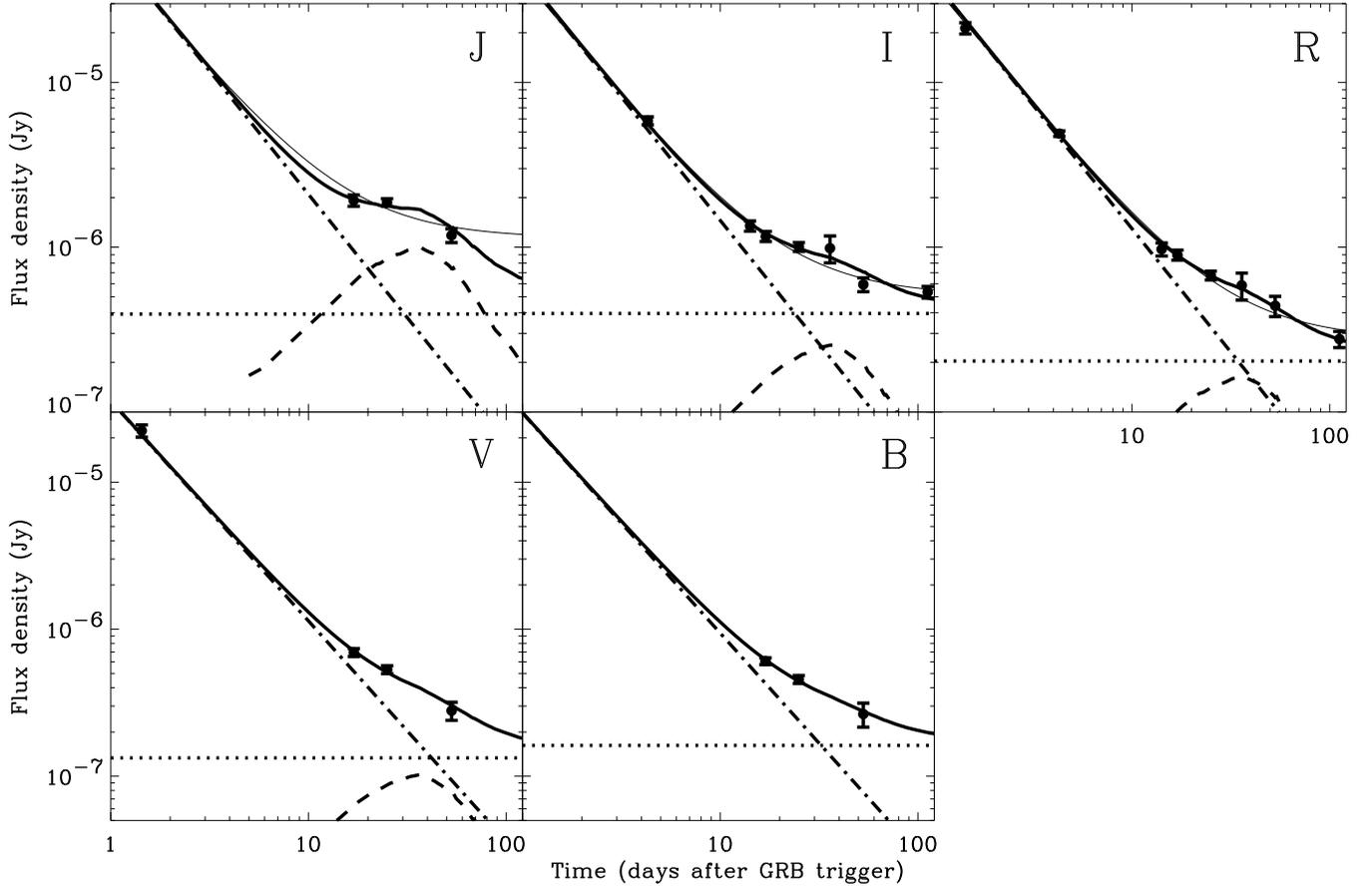,width=\textwidth}
\caption{{Lightcurves of the afterglow of GRB~000911. From top left to
bottom right, the $J$, $I$, $R$, $V$ and $B$ lightcurves are
plotted. The thick solid curves show the best fits obtained with our
three component model. The dashed, dotted and dot--dashed lines show
the SN, galaxy and ES components, respectively. The thin solid lines
in the $J$, $I$ and $R$ panels indicate the best fit for a model
comprising only the galaxy and ES (without SN). The thin line is
indistinguishable from the thick solid line for the $V$ and $B$
filters.}
\label{fig:lcur}}
\end{figure*}

The observation log and photometric measurements for the three
multifilter VLT observations are reported in Tab.~\ref{tab:log}.
Table~\ref{tab:add} shows the additional photometric measurements
taken with the Keck and MSO50 telescopes, as well as two photometric
points extracted from the VLT spectrum.  All these measurements,
together with two additional photometric points taken from the
literature (see Tab.~\ref{tab:lit}), are shown in Fig.~\ref{fig:lcur}.
The OT appears point like in all filters at all times, with the
exception of the $B$ and $V$ images at day 53, in which the OT is not
consistent with the PSF of bright stars. The faintness of the OT in
$B$ and $V$ bands at day 53, however, makes it impossible to measure a
reliable extension and ellipticity of the source.  The error bars of
photometric points were derived following the standard procedure. All
pertinent factors affecting photometric precision are taken into
account: statistical errors on the measured counts converted to
electron units, CCD read--out noise, calibration errors applying the
usual error propagation formulas. On the other hand, the photometry of
the OT was not affected by the galaxy background possibly present in
our images. In fact, we did not detect significant evidence of
deviation from the field star PSF for the OT (the only exception is
for the last $B$ and $V$ images, see above). The underlying galaxy
within the limit of our observational data seem to be consistent wih a
point--like shape perfectly superposed (i.e. within a fraction of the
FWHM of our images) to the OT. After having removed all star--like
objects in our images no diffused background emerged in the OT
region. A systematic uncertainty of 0.05 mag was added in quadrature
to all data, in order to get rid of the differences possibly
introduced by images taken with different telescopes.

\subsection{Spectroscopy}
\label{sec:obsspe}

The spectral frame reduction was again performed in a standard way
with ESO--MIDAS tools (99NOV version).  The extraction of the
scientific spectra was then obtained with IDL routines written by us.
Observing conditions were photometric.  However the closeness of the
moon to the target ($\sim 30^\circ$), the seeing variations (from
$\sim 0.8$\,arcsec to $\sim 1.8$\,arcsec) and the OT faintness made
the spectrum extraction critical. In particular it proved necessary to
rebin our spectra to the much lower resolution of 110 \AA~per pixel in
order to increase the S/N.  Two independent spectra were extracted
from different frames, one hour exposure each, and found to be
consistent with each other.  Absolute calibrations were performed
through the observation of a standard star (\object{EG21}) with the
same instrumental set--up.

The  resulting average  spectrum  is shown  by the   filled circles in
Fig.~\ref{fig:spec}. The spectrum,  despite  its very  low resolution,
was determined to  be considerably redder  than any afterglow spectrum
observed so far. A power--law fit yielded $F(\nu) \propto \nu^{-5.3
\pm 0.8}$ ($1\sigma error$, $\chi^2 =22$ for 21 degrees of freedom, 
hereafter d.o.f.).

\begin{table*}
\begin{center}
\begin{tabular}{c|c|c|c|l|c}
~~Date (UT)~~~     & ~$\Delta t$ (d)~ & Filter & Exposure time (s) & ~~Magnitude~~ 
& Seeing \\ \hline \hline
28 Sep. 2000 & 16.94         & $B$    & 2600     & $25.10 \pm 0.05$~ & 0.8$''$\\
06 Oct. 2000 & 24.98         & $B$    & 2600     & $25.41 \pm 0.06$ & 0.6$''$\\
03 Nov. 2000 & 52.95         & $B$    & 2600     & $26.0~ \pm 0.2~$ & 1.0$''$\\
28 Sep. 2000 & 16.96         & $V$    & 1500     & $24.67 \pm 0.06$ & 0.8$''$\\
06 Oct. 2000 & 25.02         & $V$    & 1500     & $24.96 \pm 0.06$ & 0.5$''$\\
03 Nov. 2000 & 52.99         & $V$    & 1500     & $25.66 \pm 0.15$ & 0.7$''$\\
28 Sep. 2000 & 16.98         & $R$    &  900     & $24.14 \pm 0.07$ & 0.8$''$\\
06 Oct. 2000 & 25.03         & $R$    &  900     & $24.45 \pm 0.06$ & 0.5$''$\\
03 Nov. 2000 & 53.00         & $R$    &  900     & $24.91 \pm 0.15$ & 0.7$''$\\
28 Sep. 2000 & 16.99         & $I$    & 1500     & $23.57 \pm 0.07$ & 0.8$''$\\
06 Oct. 2000 & 25.00         & $I$    & 1500     & $23.73 \pm 0.06$ & 0.5$''$\\
03 Nov. 2000 & 52.97         & $I$    & 1500     & $24.3~ \pm 0.1~$ & 0.7$''$\\
28 Sep. 2000 & 16.89         & $J$    & 2408     & $22.38 \pm 0.08$ & 0.7$''$\\
06 Oct. 2000 & 24.93         & $J$    & 2700     & $22.41 \pm 0.05$ & 0.5$''$\\
05 Nov. 2000 & 54.90         & $J$    & 2700     & $22.9~ \pm 0.1~$ & 0.5$''$
\end{tabular}
\caption{{Log of the VLT photometric observations of the afterglow 
of GRB~000911.}
\label{tab:log}}
\begin{tabular}{c|c|l|c|l|c}
$\Delta t$ (d) & Filter & Telescope & Exposure time (s) & Magnitude & Seeing \\ 
\hline \hline
1.435 & $R$         & MSO50    & 1200     & $20.70 \pm 0.08~$ & 2.1$''$\\
1.435 & $V$         & MSO50    & 1200     & $20.90 \pm 0.1~$  & 2.1$''$\\
~4.30 & $R$         & Keck     &  600     & $22.30 \pm 0.06$  & 0.6$''$\\
~4.31 & $I$         & Keck     &  620     & $21.82 \pm 0.06$  & 0.7$''$\\
14.12 & $R$         & Keck     & 1830     & $24.05 \pm 0.094$ & 1.1$''$ \\
14.15 & $I$         & Keck     & 1080     & $23.41 \pm 0.07$  & 0.9$''$ \\
35.92 & $R$         & VLT Spectrum &  6800     & $24.6~~ \pm 0.2~$ & 0.9$''$\\
35.92 & $I$         & VLT Spectrum &  6800     & $23.75~ \pm 0.2~$ & 0.9$''$
\end{tabular}
\caption{{Additional photometric points from Keck, MSO 50 inch telescope
and VLT spectroscopy.}
\label{tab:add}}
\begin{tabular}{c|c|c|c|l|c}
$\Delta t$ (d) & Filter & Telescope & Magnitude & Reference & Seeing \\ 
\hline \hline
111.9          & $R$   & Keck & $25.4 \pm 0.1$ & Price et al. 2001 & 0.7$''$\\
112.0          & $I$   & Keck & $24.4 \pm 0.1$ & Price et al. 2001 & 0.65$''$
\end{tabular}
\caption{{Photometric points from literature.}
\label{tab:lit}}
\end{center}
\end{table*}

\section{Data modelling}
\label{sec:mod}

We have modeled the data with a composite spectrum given by combining
an external shock synchrotron component (Meszaros \& Rees 1997) plus a
host galaxy. In addition, we examined the possible role of a supernova
component. The details of the photometry and spectral fitting are
discussed in the following subsections.

\subsection{Photometry}
\label{sec:modpho}

The photometric data were dereddened for Galactic extinction according
to the maps of Dickey and Lockman (1990) ($E_{B-V} = 0.112$) and
following extinction curves from Cardelli et al. (1989), and converted
to flux densities. Extinction in the host galaxy was not modeled
since no additional extinction was required by the data.

The lightcurve modelling  was obtained as the  sum of the  above three
contributions. First  we  considered    an external shock    afterglow
component (hereafter ES) of the form:
\begin{equation}
F(\nu,t)=A_{\rm ES} \, \nu^{-\alpha} \, t^{-2\,\alpha}.
\label{eq:exshock}
\end{equation}
This equation holds for a jet geometry after the break time (Sari,
Piran \& Halpern 1999). Such a configuration is obtained by a
broadband fitting of the GRB~000911 afterglow (Price et al. 2001).
The constant flux from a host galaxy was added in the five bands as a
free parameter ($G_B$, $G_V$, $G_R$, $G_I$ and $G_J$).

First, an ES plus galaxy model was fitted to the data.  The best fit
gave a decay slope $\alpha = 0.724\pm0.006$ (temporal slope
$\delta=1.45\pm0.012$) with $\chi^2 = 24.4$ for 18 d.o.f..  This is an
acceptable fit, with chance probability of $P\sim10\%$ to obtain a
higher $\chi^2$ value. However, the fit can be improved by adding the
SN component at the redshift $z=1.06$ of the host galaxy (see
Sect.~\ref{sec:int}).  The lightcurve of the supernova component was
obtained by spline interpolation of the data of SN1998bw (z=0.0085;
Galama et al. 1998).  Cosmological parameters
$H_0=65$~km~s$^{-1}$~Mpc$^{-1}$ and $q_0=0.5$ were adopted to compute
the flux as a function of redshift, and the time profile was stretched
by a factor $1+z$.  Following Bloom et al. (1999), the spectrum of the
supernova was analytically extended in the rest frame ultraviolet
assuming a power--law $F(\nu) \propto
\nu^{-2.8}$ for $\lambda < 3600$~\AA. SN1998bw was adopted as a
template since it is the SN with the most widely accepted association
with a GRB (GRB~980425) due to the coincident position (Galama et
al. 1998) and its peculiar relativistic expansion (Kulkarni et
al. 1998).  In addition, it has a very well sampled $UBVRI$ lightcurve
(Galama et al. 1998).

The addition of the supernova component changes slightly the spectral
slope of the ES component ($\alpha =0.748\pm 0.006$) with $\chi^2 =
13.9$ ($P\sim67\%$).  In principle, the addition of this component
does not introduce any new free parameter since the redshift is known
and the SN luminosity fixed to the value of SN1998bw. However, the
lightcurves of different SNe, even within the same class, are
different, with different peak luminosity and peak time. We hence let
the normalization of the SN component free to vary, with the redshift,
the color and the time profile held fixed.  We obtain a normalization
factor $A=0.9\pm0.3$, with $\chi^2=13.8$ for 17 d.o.f..  An F test
applied gives a statistical confidence of $99.8\%$ ($2.9\sigma$) for
the fit improvement.  Alternatively, we allowed the redshift to
vary. Interestingly, we obtain $z_{\rm SN} = 1.1$, in good agreement
with the $z=1.06$ measured spectroscopically by Price et al.  (2001)
for the host galaxy of GRB~000911.

The re--brightening component in the   lightcurve is hence  remarkably
similar  in  luminosity,   shape  and color   to   the lightcurve   of
SN1998bw. As a final test, we allowed for  a temporal shift $\Delta t$
between the SN and GRB explosions.  Keeping all the other SN parameter
fixed to the  values of SN1998bw and  the redshift $z=1.06$, we obtain
$\Delta t  = 0^{+1.5}_{-7}$~days  ($1\,\sigma$), showing  that  the SN
explosion may anticipate the GRB but only by $\sim 1$ week (see, e.g.,
the Supranova model by Vietri \& Stella 1998).

The best fit three  component model (ES plus galaxy  plus SN) is shown
by the thick solid line in Fig.~\ref{fig:lcur}.  The figure also shows
the three individual components (ES, galaxy and SN).

\begin{figure}
\psfig{file=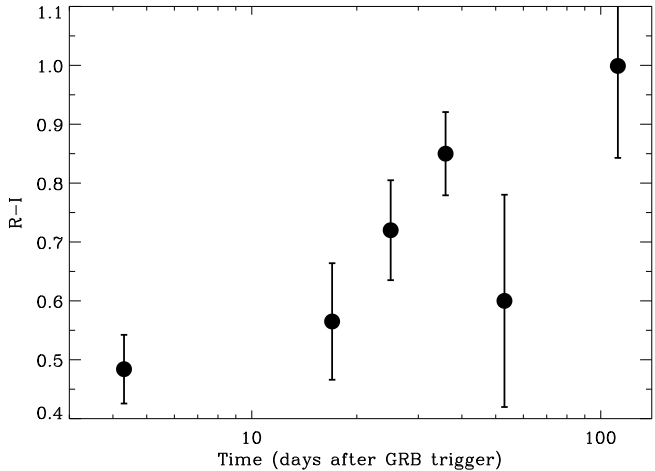,width=.48\textwidth}
\caption{{Evolution of the $R-I$ color index with the time elapsed
after the GRB event. The color shows a  significant evolution from the
value $R-I \sim 0.55$ in the first 20 days  to the final value of $R-I
\sim 0.85$ afterwards.}
\label{fig:ri}}
\end{figure}

In Fig.~\ref{fig:ri} we report the evolution of the color index $R-I$
as a function of time. The color significantly evolves from blue to
red, as time goes on.  A formal best fit with a constant color yields
$\chi^2=23$ for 5 d.o.f.. The reddening behavior can be due either to
the presence of an underlying supernova or to a red host galaxy
($R-I\sim 1$) which certainly dominates the color evolution at
late time ($t\sim100$~d). Again, the quality of the data, especially
at late times, does not allow us to unambiguously select a model.

\subsection{Spectroscopy}
\label{sec:modspe}

According to the fits described above, none of the three components
clearly dominated the emission at the time of the spectroscopic
observation ($\sim36$~days; see Fig.~\ref{fig:lcur}).  Moreover,
the wavelength range over which the spectrum has a sufficient signal
to noise to be safely extracted covers only a limited spectral range
($6000 {\rm \AA} <
\lambda < 8500 {\rm \AA}$).  For this reason, the spectrum by itself
can be modeled in different ways, all consistent with the data.

\begin{figure}
\psfig{file=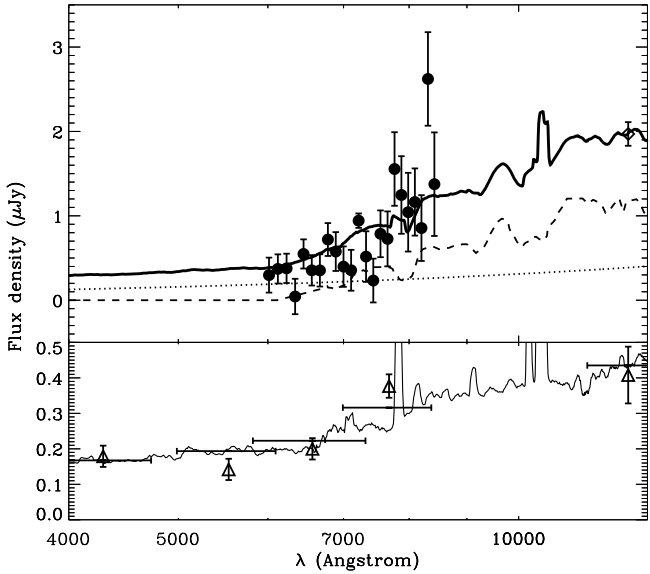,width=.48\textwidth}
\caption{{Spectrum of the OT of GRB~000911 observed 36 days after the
burst explosion (upper panel, filled dots).  The spectrum is modelled
with a SN type Ic spectrum plus a background starburst galaxy and an
ES component (see text).  The solid line shows the total spectrum
smoothed with a 110~\AA~boxcar filter, while the dashed line
represents the supernova component (SN1987M, see text). The
lower panel shows the template spectrum of the best fit galaxy
model (thin solid line). Triangles are the galaxy photometric
measurements as derived from the multiband fitting (see
sect.~\ref{sec:modpho} and the dotted horizontal lines in
Fig.~\ref{fig:lcur}). The vertical position of the horizontal
bars indicate the $BVRIJ$ filter fluxes derived from the galaxy
template; their width is equal to the full width at half maximum of
the filters.}
\label{fig:spec}}
\end{figure}

To better constrain the  models,  we added photometric information  to
the spectrum. In the upper panel  of Fig.~\ref{fig:spec}, the spectrum
(filled dots) is plotted together with the $J$ band measurement at day
25 (diamond). Note that the change in the $J$ magnitude from day 25 to
day  36   is  expected  to   be  small (see     the  first  panel   of
Fig.~\ref{fig:lcur}).    In the lower    panel,  the best fit   galaxy
components obtained with  the lightcurve fitting procedure (from  left
to right $G_B$, $G_V$, $G_R$, $G_I$ and $G_J$) are  shown for the five
photometric filters  used (triangles).   We  modeled the  ensemble  of
these data  with a galaxy   spectrum (templates from  Calzetti et  al.
1994) plus a power--law ES spectrum (parameters fixed to the best--fit
values from  the lightcurve) and  a type Ic   supernova spectrum a few
days  after the  peak (SN 1987M,  Filippenko  et  al. 1990)\footnote{A
spectrum of SN1998bw would be more suited for this comparison, however
no  spectra  of SN1998bw  at   peak extend to  wavelengths  $\lambda <
3500$~\AA, a range that is necessary for  a proper comparison with the
data of a SN at $z=1.06$.}.

The two models were fitted as follows: the total template spectrum was
fitted  to the spectral  data together with  the $J$ band measurement;
galaxy magnitudes derived from the  template galaxy were fitted to the
data in the lower panel.  For the galaxy plus  ES model, a formal best
fit   was   obtained  with     a  dust--enshrouded  starburst   galaxy
template. The fit gave $\chi^2 = 58$ for 27 d.o.f.  ($P\sim 0.05 \%$).
A better fit ($\chi^2=37$ for 27  d.o.f., $P\sim 9\%$) was obtained by
adding a  SN component, with  a moderately  dust--enshrouded starburst
galaxy ($0.11 < E_{B-V}  < 0.21$) as a template.   The fit is shown in
Fig.~\ref{fig:spec} overlaid on the data.

In   order  to obtain a   single   statistical indicator combining the
photometric and spectral information, we finally fitted simultaneously
all the available data    with the appropriate galaxy   template. This
yielded $\chi^2 = 44$ (39 d.o.f.) and $\chi^2=69$ (40 d.o.f.)  for
the models  with and without  SN component, respectively. The $\chi^2$
decrease has a statistical significance of $4\sigma$, according to the
F--test.

\section{Alternatives}
\label{sec:disc}

We detected a possible re--brightening and reddening of the OT
associated with GRB~000911 $\sim 30$ days after the GRB event. We
showed that this is consistent with the presence of an underlying SN
component.  In the following we discuss two possible alternatives.

\subsection{Dust echoes}
\label{sec:dust}

It has  been suggested that the  re--brightening detected  in the late
optical   lightcurves of  several   GRBs may   be  explained  by  dust
scattering or reprocessing   of  the burst and/or    afterglow photons
(Waxman \&  Draine 2000;  Esin \&  Blandford 2000).   The spectrum and
lightcurve of these components have been modeled in detail by Reichart
(2001).  He concludes that   the re--brightening in the lightcurve  of
GRB~970228 cannot be explained with  dust scattering and reprocessing,
because it predicts a spectral shape different from the observed one.

The same conclusion can be drawn in the case  of GRB~000911.  In fact,
the  re--brightening   due to  dust  reprocessing  has  an exponential
cut--off at wavelengths $\lambda < 2.2\,(1+z)\, \mu$m (Reichart 2001),
which lies in the Johnson $L$ filter for a  $z=1.06$ burst.  We detect
a re--brightening  in  the $R$, $I$  and  $J$  filter, and  hence this
interpretation can be  ruled out. Scattering from  dust,  on the other
hand, does not change the spectral index of the afterglow by more than
$\sim 0.5$ for intermediate--low dust opacities  (Reichart 2001).  The
spectrum we measure  at  day 36 implies a  much  larger reddening (see
Sect.~\ref{sec:obsspe}) and a  dust echo for a low--intermediate  dust
opacity $\tau_{\rm dust}  \lsim 12$ cannot  be fitted to the data even
considering the contribution from a red  host galaxy (the fit yields a
$\chi^2$ comparable  to that obtained for  the ES  plus galaxy model).
For a $\tau_{\rm dust} > 12$ dust cloud,  the re--brightening would be
so small that it would remain  undetected. The above conclusion can be
clearly further  strengthened   through  the measurement  of   the $J$
magnitude of the host galaxy, which would allow us to better constrain
the temporal and spectral shape of the re--brightening component.

\subsection{Wind--Fireball Interaction}
\label{sec:enri}

In the collapsar model for GRBs (MacFadyen \& Woosley 1999), or in any
other model involving a  massive star, the  key to obtain relativistic
motions  is the escape of  an energy--loaded fireball from the stellar
environment. A  focused low--entropy jet that  has broken free  of its
stellar cocoon is likely to arise from  a Wolf--Rayet (WR) progenitor.
WR stars are characterized by strong  stellar winds gradually shedding
most of the star envelope. The deceleration of this pre--collapse wind
by the pressure of  the surrounding medium causes  the ejected mass to
accumulate   at  a    radius $\sim10^{17} \dot{M}_{-4}   n_{0,1}^{1/2}
v_{3}^{-1/2}$ cm  (where $\dot{M}$ is the mass  loss rate in  units of
solar masses  per  year, $n_{0}$ is   the density of  the  surrounding
medium in units  of   cm$^{-3}$, $v$  the  wind velocity  in  units of
km~s$^{-1}$ and  we adopt the convention  $Q = 10^x\,Q_x$). The impact
between  the forward shock and   these high--density regions should be
observed as  a re--brightening   of  the afterglow, typically with   a
redder spectrum. Using  detailed stellar tracks  for the evolution  of
massive  stars, as described in Ramirez--Ruiz  et al. (2001), we found
that the re--brightening of GRB 000911 at $z=1.06$ can be explained by
the interaction   of a  relativistic  blast   wave expanding  into the
ambient medium expected at the end of the life of  a 40 $M_{\odot}$ (7
$M_{\odot}$ core) WR star evolving with  solar metallicity.  The shock
front  expands  within  a   $n(r)=  1.5 \times   10^{35}\,r^{-2}\,{\rm
cm}^{-1}$  stellar wind until it reaches  the density enhancement (see
Fig.  3 of  Ramirez--Ruiz et al.  2001).  If  we  consider the forward
shock emission alone (shock wave into the accumulated wind), it is not
possible to have  a   spectral slope  steeper than $\alpha   \sim 3.5$
(Ramirez--Ruiz et al. 2001).  Including   the emission of the  reverse
shock, spectral slopes as  steep as $\alpha  \sim 4$ can  be obtained,
depending on  the   radiative efficiency of  the   reverse  shock. The
spectroscopic   observation can  then   be reproduced if  some  of the
reddening is attributed to the host galaxy emission.

We have fitted this model to the photometric data only obtaining a
$\chi^2$ intermediate between the fit with a simple ES plus galaxy
model and the fit with the SN additional component. This model also
requires the addition of 4 free parameters (the Lorentz factor of the
fireball, the mass of the progenitor star, its metallicity and the
density of the surrounding medium) and the need for an enhanced
reverse shock emission to reproduced the observed spectrum. Since
the addition of more degrees of freedom does not improve the
statistics of the $\chi^2$, we consider the SN a better explanation
for the re--brightening in the GRB~000911 lightcurve.

\section{Conclusions}

We presented late time multifilter observations of the
optical transient associated to GRB~000911.  This set of observations
was designed to detect and analyze the re--brightening associated with
(some) GRB afterglows approximately one month after their explosion
(Bloom et al. 1999; Reichart 1999; Galama et al. 2000).  In addition
to photometric data, a low resolution spectrum was taken $\sim 36$
days after the burst explosion.

The lightcurve and spectrum were fitted with an external shock plus
galaxy model, with the possible addition of a supernova, similar to
SN1998bw. The addition of the SN component gives a better fit, with a
statistical significance of $4\,\sigma$.  Even though the observations
were designed to detect unambiguously the additional supernova
component, the signal to noise ratio that could be achieved was
limited, due to the combination of two effects. Firstly the burst was
at a redshift larger than in the two previous cases (GRB~970228:
$z=0.695$; GRB~980326: $z \sim 0.9$); secondly, the host galaxy
associated with GRB~000911 had a brightness comparable to that of the
SN component.  For this reason, a fourth multifilter observation at a
much longer time ($>100$~days) is mandatory in order to disentangle
the two components.  We were able to find $R$ and $I$ magnitudes at
this late times in the literature (Price et al. 2001), but the most
important measurement in the $J$ band (see the difference between the
thick and thin solid lines in the $J$ panel of Fig.~\ref{fig:lcur})
 was not available.

With the present data, a word of caution should be spent, since is not
possible to unambiguously assess the presence of the SN. However, if
future observations will allow us to better constraint the magnitude
of the host galaxy and will confirm the presence of the rebrightening,
we will be able to disentangle the SN component and to provide
simultaneous multiband SEDs at the time of our 3 VLT
observations. Such time resolved broad band SEDs will allow to better
understand the spectral evolution of the bump in the lightcurve and
hence to understand its physical origin.

\begin{acknowledgements}
We thank the ESO--ANTU service team that performed the observations.
We thank the ESO Director for the allocation of discretionary time for
the spectral observations performed on day 36. This program was
partially supported through ASI grants. SRK thanks NSF and NASA for
supporting a program of GRB investigations at Caltech.
\end{acknowledgements}

\end{document}